\documentclass[12pt]{article}

\usepackage{
mathrsfs,jheppub,amsbsy,amssymb,latexsym,amsfonts,amsmath,epsfig,multicol,bbm
}

\allowdisplaybreaks

\newcommand{\sh}{Schr\"odinger }
\newcommand{\po}{Poincar\'e }

\title{Classical integrability of Schr$\ddot{\bf o}$dinger sigma models and $q$-deformed Poincar\'e symmetry}

\author{Io Kawaguchi and Kentaroh Yoshida} 
\affiliation{Department of Physics, Kyoto University, Kyoto 606-8502, Japan}
\emailAdd{io@gauge.scphys.kyoto-u.ac.jp, kyoshida@gauge.scphys.kyoto-u.ac.jp}

\arxivnumber{1109.0872[hep-th]}

\abstract{
We discuss classical integrable structure of two-dimensional sigma models which have three-dimensional 
Schr$\ddot{\rm o}$dinger spacetimes as target spaces. The Schr$\ddot{\rm o}$dinger spacetimes 
are regarded as null-like deformations of AdS$_3$. 
The original AdS$_3$ isometry 
$SL(2,{\mathbb R})_{\rm L} \times SL(2,{\mathbb R})_{\rm R}$ is broken to 
$SL(2,{\mathbb R})_{\rm L} \times U(1)_{\rm R}$ due to the deformation. According to this symmetry, 
there are two descriptions to describe the classical dynamics of the system,  
1) the $SL(2,{\mathbb R})_{\rm L}$ description and 2) the enhanced $U(1)_{\rm R}$ description.  
In the former 1), we show that the Yangian symmetry is realized by improving the $SL(2,{\mathbb R})_{\rm L}$ 
Noether current. Then a Lax pair is constructed with the improved current and 
the classical integrability is shown by deriving the $r/s$-matrix algebra. 
In the latter 2), we find a non-local current by using a scaling limit of warped AdS$_3$ and 
that it enhances $U(1)_{\rm R}$ to a $q$-deformed Poincar\'e algebra. 
Then another Lax pair is presented and the corresponding $r/s$-matrices are also computed. 
The two descriptions are equivalent via a non-local map. 
}


\begin{document}

\maketitle

\section{Introduction}

The AdS/CFT correspondence \cite{Maldacena,GKPW} is the most concrete realization of dualities 
between gauge theories and gravitational (string) theories. In the recent progress the integrable structure 
played an important role in checking it at non-BPS region (For a comprehensive review, see \cite{review}). 
The integrability of sigma models on AdS spaces and spheres \cite{Mandal,BPR} is closely related to the integrable structure 
behind AdS/CFT and a key ingredient is that AdS spaces and spheres are represented by symmetric cosets.  
This feature was utilized to classify the coset geometries which can potentially be studied holographically \cite{Zarembo}.  
It is well known that symmetric coset sigma models are classically integrable and 
infinite-dimensional symmetries, which are often called Yangian \cite{Luscher,Drinfeld}, 
are realized (For early works and a review, see \cite{Bernard-Yangian,MacKay} and \cite{review-Yangian}). 

\medskip 

Recently, gravity duals for non-relativistic CFTs \cite{Balasubramanian,Son} are intensively 
studied in relation to the holographic condensed matter scenario. \sh spacetimes are proposed as gravity duals 
for non-relativistic CFTs \cite{Nishida} possessing non-relativistic conformal symmetry called \sh symmetry \cite{Sch1,Sch2}. 
It is an interesting issue to consider the integrability of sigma models which have the Schr{\"o}dinger spacetimes 
as target spaces\footnote{It is worth mentioning about the very recent paper \cite{recent} in which an intimate 
connection between the \sh spacetimes \cite{Anninos} and the Kerr/CFT correspondence \cite{Kerr/CFT} is discussed.}. 
The Schr{\"o}dinger spacetimes are represented by non-symmetric coset \cite{SYY}, and hence 
the standard argument for symmetric cosets is not applicable directly to Schr{\"o}dinger spacetimes. 

\medskip 

In this paper we will reveal the classical integrable structure of two-dimensional sigma models defined on 
three-dimensional Schr{\"o}dinger spacetimes, which are called ``\sh sigma models'' here. 
The metric of the \sh spacetime is  
\begin{eqnarray}
ds^{2}=L^{2}\left[d\rho^{2}-2{\rm e}^{-2\rho}dudv-C{\rm e}^{-4\rho}dv^{2}\right]\,, 
\label{angle}
\end{eqnarray}
where $C$ is a constant deformation parameter. 
When $C=0$\,, (\ref{angle}) becomes AdS$_3$ with the radius $L$\,. 
When $C\neq 0$, we can put $C=\pm 1$ with the Lorentz boost, 
\[
u \to \lambda u\,, \qquad v \to \lambda^{-1}v \qquad (\lambda:~\mbox{const.})\,. 
\]
The AdS$_3$ isometry $SO(2,2)=SL(2,{\mathbb R})_{\rm L}\times SL(2,{\mathbb R})_{\rm R}$ 
is broken to $SL(2,{\mathbb R})_{\rm L}\times U(1)_{\rm R}$ due to the non-vanishing $C$\,. 
According to $SL(2,{\mathbb R})_{\rm L}\times U(1)_{\rm R}$\,, there are two descriptions to describe 
the classical dynamics of the system,  1) the left description based on $SL(2,{\mathbb R})_{\rm L}$  
and 2) the right description based on enhanced $U(1)_{\rm R}$. 

\medskip 

In the former description 1), the Yangian symmetry is shown to be realized as a hidden symmetry by improving 
the $SL(2,{\mathbb R})_{\rm L}$ Noether current so that it satisfies the flatness condition, following \cite{KY}. 
Then a Lax pair is constructed with the improved current and the classical integrability is shown by deriving 
the $r/s$-matrix algebra. The universality class of this system is rational as is expected from the presence of Yangian. 
The argument here is quite similar to the previous works for squashed spheres and warped AdS spaces \cite{KY,KOY}. 

\medskip 

In the latter description 2), a non-local conserved current is presented 
by applying a non-local map to the improved current as in the case of squashed sphere \cite{KY3}. 
It enhances $U(1)_{\rm R}$ to a $q$-deformed Poincar\'e algebra \cite{q-Poincare,Ohn}. 
Then another Lax pair, which also leads to the classical equations of motion exactly,
is presented by taking a scaling limit from the Lax pair in the case of warped AdS$_3$ 
and the corresponding $r/s$-matrices show that the system is rational. 
This means that the left and right descriptions are 
equivalent at classical level. 
In fact, the two descriptions are equivalent via a non-local map. 

\medskip 

This paper is organized as follows. In section 2 three-dimensional \sh spacetimes are rewritten 
in terms of $SL(2,{\mathbb R})$ group element. Then the action of \sh sigma models is introduced. 
In section 3 we study the classical integrability in the left description 
based on the $SL(2,{\mathbb R})_{\rm L}$ symmetry. 
In section 4 the classical integrability is discussed in the right description based 
on enhanced $U(1)_{\rm R}$\,. 
Section 5 is devoted to conclusion and discussion. 
In Appendix A we explain the derivation of Lax pair in the right description in detail.

\section{Schr{\"o}dinger sigma models} 

\sh spacetime in any dimensions is homogeneous and can be described as a coset \cite{SYY}. 
Although the coset is non-reductive in general, an exception is the three-dimensional case 
and the coset becomes reductive. 
We are confined to this case hereafter. 

\medskip 

For the later convenient, let us introduce an $SL(2,{\mathbb R})$ group element represented by 
\begin{eqnarray}
g={\rm e}^{2vT^{+}}{\rm e}^{2\rho T^{2}}{\rm e}^{2uT^{-}}\,.
\end{eqnarray}
The $SL(2,{\mathbb R})$ generators $T^{a}~(a=0,1,2)$ are expressed in terms of the standard Pauli matrices like 
\begin{eqnarray}
&&T^{0}=\frac{i}{2}\sigma_{2}~,\qquad 
T^{1}=\frac{1}{2}\sigma_{1}~, \qquad 
T^{2}=\frac{1}{2}\sigma_{3}~, \nonumber 
\end{eqnarray}
and the light-cone notation is defined as 
\begin{eqnarray}
&&T^{\pm}=\frac{1}{\sqrt{2}}\left(T^{0}\pm T^{1}\right)\,.
\end{eqnarray}
They satisfy the relations 
\begin{eqnarray}
\left[T^{a},T^{b}\right]=\varepsilon^{ab}_{~~c}\,T^{c}\,,\qquad 
{\rm Tr}\left(T^{a}T^{b}\right)=\frac{1}{2}\gamma^{ab}\,,
\end{eqnarray}
where the anti-symmetric tensor $\varepsilon^{ab}_{~~c}$ is normalized $\varepsilon^{012}=+1$
and the metric on ${\mathbb R}^{1,2}$ is $\gamma^{ab}=(-1,+1,+1)$\,.  
The group indices are raised and lowered with $\gamma^{ab}$ and its inverse. 

\medskip 

As a result, the metric (\ref{angle}) can be rewritten as 
\begin{eqnarray}
ds^{2} &=& \frac{L^{2}}{2}\left[
{\rm Tr}\left(J^{2}\right) 
- 2C \left({\rm Tr}\left(T^{-}J\right)\right)^{2}
\right] 
\label{rewritten} \\
&=& \frac{L^{2}}{4}\left[-2J^{-}J^{+}+\left(J^{2}\right)^{2}-C\left(J^{-}\right)^{2}\right] \nonumber 
\end{eqnarray}
in terms of the left-invariant one-form $J$ defined as 
\begin{eqnarray}
&& J \equiv g^{-1}dg\,, 
\qquad J^{a}=2{\rm Tr}\left(T^{a}J\right)\,. 
\end{eqnarray}
It is easy to see that the metric (\ref{rewritten}) is invariant 
under the $SL(2,{\mathbb R})_{\rm L}\times U(1)_{\rm R}$ transformation: 
\begin{eqnarray}
g\rightarrow g^{L}\cdot g\cdot {\rm e}^{-\alpha T^{-}}\,.
\end{eqnarray}
The infinitesimal $SL(2,{\mathbb R})_{\rm L}$ and $U(1)_{\rm R}$ transformations are given by, respectively, 
\begin{eqnarray}
&& \delta^{L,a}g = \epsilon\, T^{a}\,g\,,
\label{left_tr} \\ 
&& \delta^{R,-}g = -\epsilon\, g\,T^{-}\,.
\label{right_tr}
\end{eqnarray}

\medskip 

Here it should be noted that 
AdS$_{3}$ has three kinds of anisotropic deformations, i) space-like, ii) time-like and iii) null-like deformations. 
The \sh spacetimes correspond to null-like deformations of AdS$_{3}$\,.  
The metric of space-like warped AdS$_{3}$ is realized with a deformation term on $T^1$ as  
\begin{eqnarray}
&& ds^{2}=\frac{L^{2}}{2}\left[
{\rm Tr}\left(J^{2}\right)-2\widetilde{C}\left[{\rm Tr}(T^{1}J)\right]^{2}
\right]\,.
\label{spacewarped} 
\end{eqnarray}
The metric of time-like warped AdS$_{3}$ is obtained with a deformation term on $T^0$ as 
\begin{eqnarray}
&& ds^{2}=\frac{L^{2}}{2}\left[
{\rm Tr}\left(J^{2}\right)-2\widetilde{C}\left[{\rm Tr}(T^{0}J)\right]^{2}
\right]\,.
\end{eqnarray}

\medskip 

The null-like warped AdS$_{3}$ is obtained from both space-like and time-like warped AdS$_{3}$ geometries 
by taking a scaling limit \cite{Anninos}. 
As an example, let us consider a space-like warped AdS$_{3}$ with a deformation parameter $\widetilde{C}$\,. 
The metric can be rewritten in terms of $T^{\pm}$ as  
\begin{eqnarray}
ds^{2}=\frac{L^{2}}{2}\left[
{\rm Tr}\left(J^{2}\right)-\widetilde{C}\left[{\rm Tr}(T^{+}J)\right]^{2}
+2\widetilde{C}{\rm Tr}(T^{+}J){\rm Tr}(T^{-}J)-\widetilde{C}\left[{\rm Tr}(T^{-}J)\right]^{2}
\right]\,.
\end{eqnarray}
By rescaling $T^{\pm}$ as  
\begin{eqnarray}
T^{-}\rightarrow\sqrt{\frac{2C}{\widetilde{C}}}\,T^{-}\,,\qquad 
T^{+}\rightarrow\sqrt{\frac{\widetilde{C}}{2C}}\,T^{+} 
\label{lim}
\end{eqnarray}
and taking $\widetilde{C}\rightarrow 0$ limit with $C$ fixed, the metric (\ref{rewritten}) is reproduced. 
The above argument is applicable to time-like warped AdS$_{3}$ as well.

\subsection{The action of \sh sigma models}

The action of \sh sigma models is  
\begin{eqnarray}
S&=&-\int\!\!\!\int\!\!dtdx\,\eta^{\mu\nu}\left[{\rm Tr}\left(J_{\mu}J_{\nu}\right)
-2C{\rm Tr}\left(T^{-}J_{\mu}\right){\rm Tr}\left(T^{-}J_{\nu}\right)\right]\,. 
\label{action}
\end{eqnarray}
The base space is a two-dimensional Minkowski spacetime with the coordinates $x^{\mu}=(t,x)$ 
and the metric $\eta_{\mu\nu}=(-1,+1)$. 
Although we have applications to string theory in our mind, we do not impose periodic boundary conditions and 
the Virasoro conditions for simplicity here, but the boundary condition that 
the group element variable $g(x)$ approaches a constant element very rapidly as it goes to 
spatial infinities: 
\begin{eqnarray}
g(t,x) \rightarrow g_{(\pm)}:~\mbox{const.} \quad (x \to \pm \infty)\,. 
\label{bc}
\end{eqnarray}
Thus the left-invariant current $J_{\mu}$ vanishes as it approaches spatial infinities, 
\begin{eqnarray}
J_{\mu}(t,x) \to 0 \quad (x \to \pm \infty)\,. 
\end{eqnarray}

\medskip 

The equations of motion obtained from (\ref{action}) are 
\begin{eqnarray}
\partial^{\mu}J_{\mu}-2C{\rm Tr}\left(T^{-}\partial^{\mu}J_{\mu}\right)T^{-}-2C{\rm Tr}\left(T^{-}J_{\mu}\right)\left[J^{\mu},T^{-}\right]=0\,.
\label{eom}
\end{eqnarray}
By multiplying $T^{a}$ to (\ref{eom}) and taking the trace operation, 
the $T^{a}$ component of the equations of motion can be obtained. 
The $T^{-}$ component leads to the conservation law of the $U(1)_{\rm R}$ current,  
\begin{eqnarray}
\partial^{\mu}J^{-}_{\mu}&=&0\,.
\label{eom-}
\end{eqnarray}
The $T^{2}$ and $T^{+}$ components are, respectively,  
\begin{eqnarray}
\partial^{\mu}J^{2}_{\mu}-CJ^{-}_{\mu}J^{-,\mu}&=&0\,,
\label{eom2} \\ 
\partial^{\mu}J^{+}_{\mu}-CJ^{-}_{\mu}J^{2,\mu}&=&0\,.
\label{eom+}
\end{eqnarray}
The equations of motion (\ref{eom}) are equivalent to (\ref{eom-})-(\ref{eom+}). 

\medskip 

In addition, the equations of motion (\ref{eom}) are equivalent to the conservation law of the $SL(2,{\mathbb R})_{\rm L}$ current,  
\begin{eqnarray}
\partial^{\mu}\left[gJ_{\mu}g^{-1}-2C{\rm Tr}\left(T^{-}J_{\mu}\right)gT^{-}g^{-1}\right]=0\,.
\end{eqnarray}

\medskip 

According to this observation on the equations of motion, one may expect that 
there should be two ways to describe the classical dynamics of this system. 
Indeed, this is the case. One description is based on the $SL(2,{\mathbb R})_{\rm L}$ symmetry 
and the other is on the enhanced $U(1)_{\rm R}$ symmetry.

\section{Left description based on $SL(2,{\mathbb R})_{\rm L}$}

In this section we consider the classical integrability of \sh sigma models in the left description based 
on the $SL(2,{\mathbb R})_{\rm L}$ symmetry. 
First, we show that $SL(2,{\mathbb R})_{\rm L}$ symmetry is enhanced to an infinite-dimensional symmetry, 
the $SL(2,{\mathbb R})_{\rm L}$ Yangian, by improving the $SL(2,{\mathbb R})_{\rm L}$ Noether  current. 
Then we construct Lax pair and monodromy matrix, 
and show the classical integrability by deriving the classical $r/s$-matrix algebra.

\subsection{Yangian symmetry}

The action (\ref{action}) is invariant under the $SL(2,{\mathbb R})_{\rm L}$ transformation (\ref{left_tr}). 
The corresponding conserved $SL(2,{\mathbb R})_{\rm L}$ current is 
\begin{eqnarray}
j^{L}_{\mu}=gJ_{\mu}g^{-1}-2C{\rm Tr}\left(T^{-}J_{\mu}\right)gT^{-}g^{-1}+\epsilon_{\mu\nu}\partial^{\nu}f\,.
\end{eqnarray}
The first two terms can be obtained by the Noether procedure 
but the last term is the ambiguity of the conserved current. The anti-symmetric tensor 
$\epsilon_{\mu\nu}$ on the base space is normalized 
as $\epsilon_{tx}=+1$ and $f$ is an arbitrary function. When $f$ is taken as 
\begin{eqnarray}
f=-\sqrt{C}\,gT^{-}g^{-1}\,,
\end{eqnarray}
then the current $j^{L}_{\mu}$ satisfies the flatness condition, 
\begin{eqnarray}
\epsilon^{\mu\nu}\left(\partial_{\mu}j^{L}_{\nu}-j^{L}_{\mu}j^{L}_{\nu}\right)=0\,.
\end{eqnarray}
Thus the flat and conserved $SL(2,{\mathbb R})_{\rm L}$ current has been obtained in \sh sigma models. 
This improved current enables us to construct an infinite number of conserved ``non-local'' charges, for example, 
by following the BIZZ construction \cite{BIZZ}. The first two of them are 
\begin{eqnarray}
Q^{L,a}_{(0)} &=& \int^{\infty}_{-\infty}\!\!\!dx\, j^{L,a}_{t}(x)\,, \nonumber \\
Q^{L,a}_{(1)} &=& \frac{1}{4}\int^{\infty}_{-\infty}\int^{\infty}_{-\infty}\!\!\!dxdy\,
\epsilon(x-y)\varepsilon^{a}_{~~bc}\,j^{L,b}_{t}(x)j^{L,c}_{t}(y)
- \int^{\infty}_{-\infty}\!\!\!dx\, j^{L,a}_{x}(x)\,,
\end{eqnarray}
where $\epsilon(x-y) \equiv \theta(x-y) - \theta(y-x)$ and $\theta(x)$ is a step function. 

\medskip 

The next is to compute the Poisson brackets of the charges. 
For this purpose, the current algebra of the flat and conserved $SL(2,{\mathbb R})_{\rm L}$ current is needed. 
It can be computed by evaluating the standard Poisson brackets of the dynamical variables contained in the classical action 
and is written down in terms of the component of the current, 
e.g $j^{L,a}_{\mu}=2{\rm Tr}(T^{a}j^{L}_{\mu})$\,, 
\begin{eqnarray}
\left\{j_{t}^{L,a}(x),j_{t}^{L,b}(y)\right\}_{\rm P}&=&\varepsilon^{ab}_{~~c}\,j_{t}^{L,c}(x)\delta(x-y)\,, \nonumber \\
\left\{j_{t}^{L,a}(x),j_{x}^{L,b}(y)\right\}_{\rm P}&=&\varepsilon^{ab}_{~~c}\,j_{x}^{L,c}(x)\delta(x-y)
+\gamma^{ab}\partial_{x}\delta(x-y)\,, \label{current-AdS} \\
\left\{j_{x}^{L,a}(x),j_{x}^{L,b}(y)\right\}_{\rm P}&=&0\,. \nonumber 
\end{eqnarray}
Note that the current algebra does not contain $C$ explicitly 
and is exactly the same as the $SL(2,{\mathbb R})_{\rm L}$ current algebra in sigma models defined on undeformed AdS$_{3}$\,. 
It may be natural if one notices that $C$ can be absorbed into the normalization of the generators by a simple rescaling, 
\begin{eqnarray}
T^{-}\rightarrow\frac{1}{\sqrt{|C|}}\,T^{-}\,,\qquad T^{+}\rightarrow\sqrt{|C|}\,T^{+}\,.
\end{eqnarray}
Thus the Poisson brackets of the conserved charges are also the same as the AdS$_{3}$ case, 
\begin{eqnarray}
\left\{Q_{(0)}^{L,a},Q_{(0)}^{L,b}\right\}_{\rm P}&=&\varepsilon^{ab}_{~~c}\,Q_{(0)}^{L,c}\,, \nonumber \\
\left\{Q_{(1)}^{L,a},Q_{(0)}^{L,b}\right\}_{\rm P}&=&\varepsilon^{ab}_{~~c}\,Q_{(1)}^{L,c}\,, \nonumber \\
\left\{Q_{(1)}^{L,a},Q_{(1)}^{L,b}\right\}_{\rm P}&=&\varepsilon^{ab}_{~~c}\left[Q_{(2)}^{L,c}
+\frac{1}{12}\left(Q_{(0)}^{L}\right)^{2}Q_{(0)}^{L,c}\right]\,,
\end{eqnarray}
and therefore an infinite number of conserved charges satisfy the $SL(2,{\mathbb R})_{\rm L}$ Yangian algebra, as a matter of course.

\medskip 

There is another way to reproduce the current algebra (\ref{current-AdS}). The flat conserved $SL(2,{\mathbb R})_{\rm L}$ current 
is found in sigma models on space-like warped AdS$_{3}$ (\ref{spacewarped}) via a double Wick rotation as discussed in \cite{KY}. 
The current algebra in the case of space-like warped AdS$_3$ is 
\begin{eqnarray}
\left\{j_{t}^{L,a}(x),j_{t}^{L,b}(y)\right\}_{\rm P}&=&\varepsilon^{ab}_{~~c}\,j_{t}^{L,c}(x)\delta(x-y)\,, \nonumber \\
\left\{j_{t}^{L,a}(x),j_{x}^{L,b}(y)\right\}_{\rm P}&=&\varepsilon^{ab}_{~~c}\,j_{x}^{L,c}(x)\delta(x-y)+(1+\widetilde{C})\gamma^{ab}\partial_{x}\delta(x-y)\,, \nonumber \\
\left\{j_{x}^{L,a}(x),j_{x}^{L,b}(y)\right\}_{\rm P}&=&-\widetilde{C}\,\varepsilon^{ab}_{~~c}\,j_{t}^{L,c}(x)\delta(x-y)\,. \nonumber
\end{eqnarray}
The rescaling (\ref{lim}) does not change the algebra at all. Thus, by taking the limit $\widetilde{C}\rightarrow 0$\,, 
the current algebra (\ref{current-AdS}) is reproduced.

\subsection{Lax pair, monodromy matrix, $r/s$-matrices}

The improved $SL(2,{\mathbb R})_{\rm L}$ current enables us to construct a Lax pair,
\begin{eqnarray}
L^{L}_{t}(x;\lambda) = \frac{1}{1-\lambda^{2}}\left[j^{L}_{t}(x)-\lambda j^{L}_{x}(x)\right]\,, \quad 
L^{L}_{x}(x;\lambda) = \frac{1}{1-\lambda^{2}}\left[j^{L}_{x}(x)-\lambda j^{L}_{t}(x)\right]\,.
\label{leftlax}
\end{eqnarray}
Here $\lambda$ is a spectral parameter. 
The commutation relation 
\begin{eqnarray}
\left[\partial_{t}-L^{L}_{t}(\lambda),\partial_{x}-L^{L}_{x}(\lambda)\right]=0
\end{eqnarray}
reproduces the conservation law of the improved current (equivalently equations of motion) and the flat condition. 

\medskip 

Now let us introduce the monodromy matrix $U^{L}(\lambda)$ defined as 
\begin{eqnarray}
U^{L}(\lambda) \equiv {\rm P}\exp{\left[\int^{\infty}_{-\infty}\!\!\!dxL^{L}_{x}(x;\lambda)\right]}\,. 
\label{leftU}
\end{eqnarray}
The symbol P denotes the path ordering. It is easy to see that the monodromy matrix is conserved, 
\begin{eqnarray}
\frac{d}{dt}U^{L}(\lambda)=0\,.
\end{eqnarray}
Thus it can be regarded as a generating function of conserved charges. 
The expression of the conserved quantities depend on the expansion point. 
For example, when the monodromy matrix is expanded around $\lambda = \infty$\,, 
the Yangian charges we have discussed so far are reproduced. 

\medskip 

The Poisson bracket of $L^{L,a}_{x}(x;\lambda)$ is evaluated as 
\begin{eqnarray}
\left\{L^{L,a}_{x}(x;\lambda), L^{L,b}_{x}(y;\mu)\right\}_{\rm P}&=&\frac{1}{\lambda-\mu}\varepsilon^{ab}_{~~c}\left[\frac{\mu^{2}}{1-\mu^{2}}L^{c}_{x}(x;\lambda)-\frac{\lambda^{2}}{1-\lambda^{2}}L^{L,c}_{x}(x;\mu)\right]\delta(x-y) \nonumber \\
&-&\frac{\lambda+\mu}{(1-\lambda^{2})(1-\mu^{2})}\gamma^{ab}\partial_{x}\delta(x-y)\,. \label{3.12}
\end{eqnarray}
With the tensor product notation, it can be rewritten as follows:
\begin{eqnarray}
\left\{L^{L}_{x}(x;\lambda),\otimes L^{L}_{x}(y;\mu)\right\}_{\rm P}
&=& \left[r^{L}(\lambda,\mu),L^{L}_{x}(x;\mu)\otimes 1+1\otimes L^{L}_{x}(x;\mu)\right]\delta(x-y) \nonumber \\
&& -\left[s^{L}(\lambda,\mu),L^{L}_{x}(x;\mu)\otimes 1-1\otimes L^{L}_{x}(x;\mu)\right]\delta(x-y) \nonumber \\
&& -2s^{L}(\lambda,\mu)\partial_{x}\delta(x-y)\,. 
\end{eqnarray}
Here we have introduced classical $r$-matrix $r^{L}(\lambda,\mu)$ and $s$-matrix $s^{L}(\lambda,\mu)$ \cite{Maillet}\,, 
respectively, defined as 
\begin{eqnarray}
&&r^{L}(\lambda,\mu) \equiv \frac{1}{2\left(\lambda-\mu\right)}\left(\frac{\mu^{2}}{1-\mu^{2}}+\frac{\lambda^{2}}{1-\lambda^{2}}\right)\left(-T^{+}\otimes T^{-}-T^{-}\otimes T^{+}+T^{2}\otimes T^{2}\right)\,, \nonumber \\
&&s^{L}(\lambda,\mu) \equiv \frac{\lambda+\mu}{2\left(1-\lambda^{2}\right)\left(1-\mu^{2}\right)}\left(-T^{+}\otimes T^{-}-T^{-}\otimes T^{+}+T^{2}\otimes T^{2}\right)\,.
\end{eqnarray}
It is easy to show the extended classical Yang-Baxter equation is satisfied,
\begin{eqnarray}
&&\left[(r+s)^{L}_{13}(\lambda,\nu),(r-s)^{L}_{12}(\lambda,\mu)\right]
+\left[(r+s)^{L}_{23}(\mu,\nu),(r+s)^{L}_{12}(\lambda,\mu)\right] \nonumber \\
&& \qquad +\left[(r+s)^{L}_{23}(\mu,\nu),(r+s)^{L}_{13}(\lambda,\nu)\right]=0\,, 
\end{eqnarray}
where the subscripts denote the vector spaces on which the $r$- and $s$-matrices act. 
Thus the classical integrability has been shown in the left description.

\section{Right description based on enhanced $U(1)_{\rm R}$}

In this section, we describe the classical dynamics of \sh sigma models in the right description 
based on the enhanced $U(1)_{\rm R}$ symmetry. We will show that the broken components of $SL(2,{\mathbb R})_{\rm R}$ 
are realized as non-local symmetries\footnote{For an earlier argument on non-locality of the right symmetry, 
based on a T-duality, see \cite{ORU}.}. The algebra of the corresponding conserved charges is found to be $q$-deformed 
two-dimensional \po algebra \cite{q-Poincare,Ohn}. In addition, the Lax pair related to the $q$-deformed \po symmetry is 
constructed. The resulting classical $r/s$-matrices also satisfies the classical Yang-Baxter equation. 

\subsection{$q$-deformed \po symmetry}

We first show that $q$-deformed \po symmetry is realized as a non-local symmetry in \sh sigma models\,. 

\medskip 

Now the $SL(2,{\mathbb,R})_{\rm R}$ symmetry of the original AdS$_3$ is broken to $U(1)_{\rm R}$ due to the deformation. 
This is generated by $T^{-}$ as in (\ref{right_tr}) and the conserved $U(1)_{\rm R}$ current is 
\begin{eqnarray}
j^{R,-}_{\mu}&=&-2{\rm Tr}(T^{-}J_{\mu}) = -J^{-}_{\mu}\,. \nonumber 
\end{eqnarray}
In contrast to the $T^{-}$ component, the other components generated by $T^{2}$ and $T^{+}$ are not the isometry of the \sh spacetime. 
However, an important observation is that there should be a non-local symmetry 
even in the case of \sh spacetime, in analogy with our previous work \cite{KY3} on squashed spheres and warped AdS spaces.  
By following the procedure in \cite{KY3} and applying a simple non-local map\footnote{This map is analogous to the Seiberg-Witten map \cite{SW}. } to the flat and conserved 
$SL(2,{\mathbb R})_{\rm L}$ current,  the non-local current is given by
\begin{eqnarray}
j^{R,2}_{\mu} &=& -2{\rm e}^{\sqrt{C}\,\chi}{\rm Tr}\left(T^{2}g^{-1}j^{L}_{\mu}g\right)\,, \nonumber \\
j^{R,+}_{\mu} &=& -2{\rm e}^{\sqrt{C}\,\chi}{\rm Tr}\left(T^{+}g^{-1}j^{L}_{\mu}g\right)\,, \\
j^{R,-}_{\mu} &=& -2{\rm Tr}\left(T^{-}g^{-1}j^{L}_{\mu}g\right)\,. \nonumber 
\end{eqnarray}
The non-locality comes through the non-local field $\chi$ defined as  
\begin{eqnarray}
\chi(x) \equiv -\frac{1}{2}\int^{\infty}_{-\infty}\!\!\!dy\,\epsilon(x-y)\,j^{R,-}_{t}(y)\,. 
\end{eqnarray}
The boundary conditions (\ref{bc}) ensure the convergence of the integral for an arbitrary value of $x$\,. 

\medskip 

The $(2,+)$-components of the non-local current are explicitly written down as  
\begin{eqnarray}
j^{R,2}_{\mu} &=& -2{\rm e}^{\sqrt{C}\,\chi} \left[{\rm Tr}(T^{2}J_{\mu}) 
+ \sqrt{C}\,\epsilon_{\mu\nu}{\rm Tr}(T^{-}J^{\nu})\right] \nonumber \\
&=& - {\rm e}^{\sqrt{C}\,\chi} \left(J^{2}_{\mu} + \sqrt{C}\,\epsilon_{\mu\nu}J^{-,\nu}\right)\,, \nonumber \\
j^{R,+}_{\mu} &=& -2{\rm e}^{\sqrt{C}\,\chi} \left[{\rm Tr}(T^{+}J_{\mu}) + C{\rm Tr}(T^{-}J_{\mu}) 
+ \sqrt{C}\,\epsilon_{\mu\nu}{\rm Tr}(T^{2}J^{\nu})\right] \nonumber \\
&=& - {\rm e}^{\sqrt{C}\,\chi}\left(J^{+}_{\mu}+CJ^{-}_{\mu} 
+ \sqrt{C}\,\epsilon_{\mu\nu}J^{2,\nu}\right)\,.
\end{eqnarray}
Note that $\chi$ satisfies the following relation, 
\begin{eqnarray}
\epsilon_{\mu\nu}\partial^{\nu}\chi=-j^{R,-}_{\mu}\,.
\label{deriv_chi}
\end{eqnarray}
To show the conservation of the non-local currents, 
we need to use the relations (\ref{eom-})-(\ref{eom+}) and (\ref{deriv_chi}). 

\medskip 

The standard Noether charge 
\begin{eqnarray}
Q^{R,-}=\int^{\infty}_{-\infty}\!\!\!dx\,j^{R,-}_{t}(x)
\end{eqnarray}
generates the right action of $U(1)_{\rm R}$,
\begin{eqnarray}
\delta^{R,-}g=\left\{g,Q^{R,-}\right\}_{\rm P}=-gT^{-}\,.
\end{eqnarray} 
Similarly, non-local charges 
\begin{eqnarray}
Q^{R,2}=\int^{\infty}_{-\infty}\!\!\!dx\, j^{R,2}_{t}(x)\,,\qquad Q^{R,+}=\int^{\infty}_{-\infty}\!\!\!dx\, j^{R,+}_{t}(x) 
\nonumber 
\end{eqnarray}
generate non-local transformations, 
\begin{eqnarray}
\delta^{R,2}g&=&\left\{g,Q^{R,2}\right\}_{\rm P} 
=-g\left[T^{2}{\rm e}^{\sqrt{C}\chi}-\sqrt{C}T^{-}\xi^{2}\right]\,, \nonumber \\
\delta^{R,+}g&=&\left\{g,Q^{R,+}\right\}_{\rm P} 
=-g\left[T^{+}{\rm e}^{\sqrt{C}\chi}-\sqrt{C}T^{-}\xi^{+}\right]\,.
\label{nonlocal_tr}
\end{eqnarray}
Here we have introduced new non-local fields,  
\begin{eqnarray}
\xi^{2}(x)=-\frac{1}{2}\int^{\infty}_{-\infty}\!\!\!dy\, \epsilon(x-y)\, j^{R,2}_{t}(y)\,,\quad
\xi^{+}(x)=-\frac{1}{2}\int^{\infty}_{-\infty}\!\!\!dy\, \epsilon(x-y)\, j^{R,+}_{t}(y)\,. \nonumber 
\end{eqnarray}
Note that $\xi^{2}$ and $\xi^{+}$ are well defined under the boundary conditions (\ref{bc}). 
We can directly check that the action (\ref{action}) is invariant under the transformations (\ref{nonlocal_tr}). 
To show the invariance, we need to use the equations of motion (\ref{eom}) and thus 
the non-local transformations (\ref{nonlocal_tr}) are the on-shell symmetry. 

\medskip 

The Poisson brackets of $j^{R}_{t}(x)$ are
\begin{eqnarray}
\left\{j^{R,+}_{t}(x),j^{R,-}_{t}(y)\right\}_{\rm P}&=&-j^{R,2}_{t}(x)\delta(x-y)\,, \nonumber \\
\left\{j^{R,+}_{t}(x),j^{R,2}_{t}(y)\right\}_{\rm P}&=&
-{\rm e}^{\sqrt{C}\,\chi}j^{R,+}_{t}(x)\delta(x-y)
-\frac{\sqrt{C}}{2}\epsilon(x-y)j^{R,+}_{t}(x){\rm e}^{\sqrt{C}\,\chi}j^{R,-}_{t}(y) \nonumber \\
&=&\frac{1}{2}j^{R,+}_{t}(x)\partial_{y}\left[{\rm e}^{\sqrt{C}\,\chi(y)}\epsilon(x-y)\right]\,, \nonumber \\
\left\{j^{R,-}_{t}(x),j^{R,2}_{t}(y)\right\}_{\rm P}&=&{\rm e}^{\sqrt{C}\,\chi}j^{R,-}_{t}(x)\delta(x-y) \nonumber \\
&=&-\frac{1}{\sqrt{C}}\partial_{x}\left[{\rm e}^{\sqrt{C}\,\chi(x)}\right]\delta(x-y)\,.
\label{rightcurrentalg}
\end{eqnarray}
With (\ref{rightcurrentalg}) and the relations 
\begin{eqnarray}
\chi(\pm\infty)=\mp \frac{1}{2} Q^{R,-}\,,
\label{inftychi}
\end{eqnarray}
the Poisson brackets of $Q^{R,a}$ are evaluated as 
\begin{eqnarray}
\left\{Q^{R,+},Q^{R,-}\right\}_{\rm P}&=&-Q^{R,2}\,, \nonumber \\
\left\{Q^{R,+},Q^{R,2}\right\}_{\rm P}&=&-Q^{R,+}\cosh\left(\frac{\sqrt{C}}{2}Q^{R,-}\right)\,, \nonumber \\
\left\{Q^{R,-},Q^{R,2}\right\}_{\rm P}&=&\frac{2}{\sqrt{C}}\sinh\left(\frac{\sqrt{C}}{2}Q^{R,-}\right)\,.
\end{eqnarray}
In the $C\rightarrow 0$ limit, this algebra becomes the $SL(2,{\mathbb R})$ algebra. 

\medskip 

In order to get a familiar expression, let us rescale $Q^{R,+}$ as 
\begin{eqnarray}
Q^{R,+}\rightarrow \frac{\sqrt{C}}{2}Q^{R,+}\,.
\end{eqnarray}
Then the algebra is rewritten as 
\begin{eqnarray}
\left\{Q^{R,+},Q^{R,-}\right\}_{\rm P}&=&-\frac{\sqrt{C}}{2}Q^{R,2}\,, \nonumber \\
\left\{Q^{R,+},Q^{R,2}\right\}_{\rm P}&=&-Q^{R,+}\cosh\left(\frac{\sqrt{C}}{2}Q^{R,-}\right)\,, \nonumber \\
\left\{Q^{R,-},Q^{R,2}\right\}_{\rm P}&=&\frac{2}{\sqrt{C}}\sinh\left(\frac{\sqrt{C}}{2}Q^{R,-}\right)
\label{q_deformed_poincare}
\end{eqnarray}
and this algebra is known as a $q$-deformed \po algebra \cite{q-Poincare,Ohn}.
A two-dimensional \po algebra is reproduced from this expression in the $C\rightarrow 0$ limit.

\subsection{Lax pair, monodromy matrix. $r/s$-matrices}

Let us next consider a Lax pair in the right description.  
The following Lax pair, 
\begin{eqnarray}
&& L^{R}_{t}(x;\lambda)=\frac{1}{2}\left[L^{R}_{+}(x;\lambda)+L^{R}_{-}(x;\lambda)\right]\,, \nonumber \\ 
&& L^{R}_{x}(x;\lambda)=\frac{1}{2}\left[L^{R}_{+}(x;\lambda)-L^{R}_{-}(x;\lambda)\right]\,, \label{rightlax} \\
&& L^{R}_{\pm}(x;\lambda)=-\frac{1}{1\pm\lambda}\left\{-T^{+}J^{-}_{\pm}-T^{-}\left[J^{+}_{\pm} 
\mp C\left(\lambda\pm\frac{\lambda^{2}}{2}\right)J^{-}_{\pm}\right]+T^{2}J^{2}_{\pm}\right\}\,, \nonumber \\
&& J_{\pm}=J_{t}\pm J_{x} \nonumber 
\end{eqnarray}
leads to the equations of motion (\ref{eom}). This Lax pair can be reproduced by taking an appropriate scaling 
limit of the Lax pair in the warped AdS$_{3}$ cases, as explained in detail in Appendix. 

\medskip 

It is a simple practice to show the commutation relation 
\begin{eqnarray}
\left[\partial_{t}-L^{R}_{t}(\lambda),\partial_{x}-L^{R}_{x}(\lambda)\right]=0
\end{eqnarray}
leads to the equations of motion and the monodromy matrix defined as  
\begin{eqnarray}
U^{R}(\lambda) \equiv {\rm P}\exp{\left[\int^{\infty}_{-\infty}\!\!\!dxL^{R}_{x}(x;\lambda)\right]}
\end{eqnarray}
is conserved: 
\begin{eqnarray}
\frac{d}{dt}U^{R}(\lambda)=0\,.
\end{eqnarray}

\medskip 

The Poisson brackets of the spatial components of Lax pair are given by 
\begin{eqnarray}
\left\{L^{R,-}_{x}(x;\lambda),L^{R,+}_{x}(y;\mu)\right\}_{\rm P}&=&\frac{1}{\lambda-\mu}\left[\frac{\mu^{2}}{1-\mu^{2}}L^{R,2}_{x}(x;\lambda)-\frac{\lambda^{2}}{1-\lambda^{2}}L^{R,2}_{x}(x;\mu)\right]\delta(x-y) \nonumber \\
&&+\frac{\lambda+\mu}{\left(1-\lambda^{2}\right)\left(1-\mu^{2}\right)}\partial_{x}\delta(x-y)\,, \nonumber \\
\left\{L^{R,-}_{x}(x;\lambda),L^{R,2}_{x}(y;\mu)\right\}_{\rm P}&=&\frac{1}{\lambda-\mu}\left[\frac{\mu^{2}}{1-\mu^{2}}L^{R,-}_{x}(x;\lambda)-\frac{\lambda^{2}}{1-\lambda^{2}}L^{R,-}_{x}(x;\mu)\right]\delta(x-y)\,, \nonumber \\
\left\{L^{R,+}_{x}(x;\lambda),L^{R,2}_{x}(y;\mu)\right\}_{\rm P}&=&\frac{1}{\lambda-\mu}\left[-\frac{\mu^{2}}{1-\mu^{2}}L^{R,+}_{x}(x;\lambda)+\frac{\lambda^{2}}{1-\lambda^{2}}L^{R,+}_{x}(x;\mu)\right]\delta(x-y) \nonumber \\
&&+\frac{C}{2}\left(\lambda-\mu\right)\frac{\lambda^{2}}{1-\lambda^{2}}L^{R,-}_{x}(x;\mu)\delta(x-y)\,, \nonumber \\
\left\{L^{R,-}_{x}(x;\lambda),L^{R,-}_{x}(y;\mu)\right\}_{\rm P}&=&0\,, \nonumber \\
\left\{L^{R,+}_{x}(x;\lambda),L^{R,+}_{x}(y;\mu)\right\}_{\rm P}&=&\frac{C}{2}\left(\lambda-\mu\right)\left[\frac{\mu^{2}}{1-\mu^{2}}L^{R,2}_{x}(x;\lambda)+\frac{\lambda^{2}}{1-\lambda^{2}}L^{R,2}_{x}(x;\mu)\right]\delta(x-y) \nonumber \\
&&-\frac{C}{2}\frac{\left(\lambda+\mu\right)\left(\lambda-\mu\right)^{2}}{\left(1-\lambda^{2}\right)\left(1-\mu^{2}\right)}\partial_{x}\delta(x-y)\,, \nonumber \\
\left\{L^{R,2}_{x}(x;\lambda),L^{R,2}_{x}(y;\mu)\right\}_{\rm P}&=&-\frac{\lambda+\mu}{\left(1-\lambda^{2}\right)\left(1-\mu^{2}\right)}\partial_{x}\delta(x-y)\,. \nonumber 
\end{eqnarray}
With the tensor product notation, it is possible to rewrite the above brackets into a simple form,  
\begin{eqnarray}
&&\left\{L^{R}_{x}(x;\lambda),\otimes L^{R}_{x}(y;\mu)\right\}_{\rm P}
=\left[r^{R}(\lambda,\mu),L^{R}_{x}(x;\mu)\otimes 1+1\otimes L^{R}_{x}(x;\mu)\right]\delta(x-y) \nonumber \\
&&\qquad\qquad\qquad\qquad\qquad~-\left[s^{R}(\lambda,\mu),L^{R}_{x}(x;\mu)\otimes 1-1\otimes L^{R}_{x}(x;\mu)\right]\delta(x-y) \nonumber \\
&&\qquad\qquad\qquad\qquad\qquad~-2s^{R}(\lambda,\mu)\partial_{x}\delta(x-y)\,, \nonumber 
\end{eqnarray}
where we have introduced the $r$- and $s$-matrices defined as, respectively, 
\begin{eqnarray}
r^{R}(\lambda,\mu) &=& \frac{1}{2\left(\lambda-\mu\right)}\left(\frac{\mu^{2}}{1-\mu^{2}}+\frac{\lambda^{2}}{1-\lambda^{2}}\right)\left(-T^{+}\otimes T^{-}-T^{-}\otimes T^{+}+T^{2}\otimes T^{2}\right) \nonumber \\
&&+\frac{C}{4}\left(\lambda-\mu\right)\left(\frac{\mu^{2}}{1-\mu^{2}}+\frac{\lambda^{2}}{1-\lambda^{2}}\right)T^{-}\otimes T^{-}\,, \nonumber \\
s^{R}(\lambda,\mu) &=& \frac{\lambda+\mu}{2\left(1-\lambda^{2}\right)\left(1-\mu^{2}\right)}\left(-T^{+}\otimes T^{-}-T^{-}\otimes T^{+}+T^{2}\otimes T^{2}\right) \nonumber \\
&&+ \frac{C\left(\lambda+\mu\right)\left(\lambda-\mu\right)^{2}}{4\left(1-\lambda^{2}\right)\left(1-\mu^{2}\right)}T^{-}\otimes T^{-}\,.
\end{eqnarray}
It is easy to show that the extended classical Yang-Baxter equation is satisfied,  
\begin{eqnarray}
&&\left[(r+s)^{R}_{13}(\lambda,\nu),(r-s)^{R}_{12}(\lambda,\mu)\right]
+\left[(r+s)^{R}_{23}(\mu,\nu),(r+s)^{R}_{12}(\lambda,\mu)\right] \nonumber \\
&&\qquad +\left[(r+s)^{R}_{23}(\mu,\nu),(r+s)^{R}_{13}(\lambda,\nu)\right]=0\,.
\end{eqnarray}
Thus the classical integrability has been shown also in the right description.

\section{Conclusion and Discussion}

We have discussed the classical integrable structure of \sh sigma models.  
Its classical dynamics can be described by the two descriptions, 
1) the left description based on $SL(2,{\mathbb R})_{\rm L}$ and 2) the right description based on enhanced $U(1)_{\rm R}$\,. 

\medskip 

The left description is based on the $SL(2,{\mathbb R})_{\rm L}$ symmetry. 
The symmetry is enhanced to the Yangian symmetry. 
To construct the Yangian charges 
the flat and conserved $SL(2,{\mathbb R})_{\rm L}$ current is used. 
By using the current, 
one can also construct the Lax pair. 
This Lax pair leads to the rational classical $r/s$-matrix algebra. 

\medskip 

The right description is based on the enhanced $U(1)_{\rm R}$ symmetry. 
We have shown that a non-local symmetry is realized and it enhances $U(1)_{\rm R}$ to a $q$-deformed \po symmetry.  
The Lax pair and monodromy matrix concerning the hidden symmetry have also been constructed 
by taking a scaling limit of the Lax pair in sigma models defined on warped AdS$_{3}$ geometries. 
The classical $r/s$-matrices explicitly depend on the value of $C$\,, 
but nevertheless those satisfy the classical Yang-Baxter equation.

\medskip 

The two descriptions are equivalent via a non-local map. In fact, as in the case of squashed S$^{3}$ and warped AdS$_3$ \cite{KY3},   
one can figure out the map between the improved $SL(2,{\mathbb R})_{\rm L}$ current 
and the non-local current concerning the enhanced $U(1)_{\rm R}$ as follows:  
\begin{eqnarray}
j^{R,-}_{\mu} &=& -2{\rm Tr}\left(T^{-}g^{-1}j^{L}_{\mu}g\right)\,, \qquad 
j^{R,2}_{\mu} = -2{\rm e}^{\sqrt{C}\chi}{\rm Tr}\left(T^{2}g^{-1}j^{L}_{\mu}g\right)\,, \nonumber \\
j^{R,+}_{\mu} &=& -2{\rm e}^{\sqrt{C}\chi}{\rm Tr}\left(T^{+}g^{-1}j^{L}_{\mu}g\right)\,.
\end{eqnarray}
Note that this is the map within the universality class of rational type, while 
there exists a map between the rational and the trigonometric in cases of squashed S$^{3}$ and warped AdS$_3$\,. 

\medskip 

One of the next steps is to construct and solve the corresponding lattice statistical model, 
which should be called ``null-deformed spin chain models (XXN model)''. 
It seems non-diagonalizable and we are not sure whether it is well defined or not. 
It may be interesting to consider how the Bethe ansatz equations are modified in this system, 
for example, by taking a scaling limit. Indeed, quantum solutions for squashed spheres are already known 
\cite{quantum1,quantum2,quantum3}
and so it would not be difficult to extend them to the warped AdS$_3$ cases. 
The $q$-deformed \po symmetry should be realized in the ``XXN model" 
and hence the S-matrix should get some constraint by the $q$-deformed \po symmetry. 
It would also be nice to analyze the \sh sigma models at quantum level following \cite{Bernard} 
as another direction. 

\medskip 

It would be a challenging problem to try to extend the present argument to higher dimensional cases. 
The coset does not satisfy the reductive condition any more, hence it would be difficult to follow the present analysis 
completely. However, since higher-dimensional \sh algebras always contain $SL(2,{\mathbb R})_{\rm L} \times U(1)_{\rm R}$ 
as a subalgebra, we may expect to use he classical integrability discussed here 
to describe, at least,   
the motions restricted to a subspace described as a three-dimensional \sh spacetime. 

\medskip 

It is also interesting to consider the relation of our result to the recent progress on the Kerr/CFT correspondence 
\cite{recent} from the view point of integrability.

\subsection*{Acknowledgments}

We would like to thank Takashi Okada, Domenico Orlando and Akihiro Tsuchiya 
for illuminating discussions. The work of IK was supported by the Japan Society for the Promotion of Science (JSPS). 
The work of KY was supported by the scientific grants from the Ministry of Education, Culture, Sports, Science 
and Technology (MEXT) of Japan (No.\,22740160). This work was also supported in part by the Grant-in-Aid 
for the Global COE Program ``The Next Generation of Physics, Spun from Universality and Emergence'' 
from MEXT, Japan.

\appendix

\section*{Appendix}

\section{Derivation of Lax pair in the right description}

The derivation of the Lax pair in the right-description (\ref{rightlax}) is a bit complicated 
and hence it is explained in detail here.

\medskip 

We begin with the action of sigma models defined on space-like warped AdS$_{3}$ spaces,   
\begin{eqnarray}
S&=&-\int\!\!\!\int\!\!dtdx\,\eta^{\mu\nu}\left[{\rm Tr}\left(J_{\mu}J_{\nu}\right)
-2\widetilde{C}\,{\rm Tr}\left(T^{1}J_{\mu}\right){\rm Tr}\left(T^{1}J_{\nu}\right)\right]\,. 
\label{space_action}
\end{eqnarray}
The classical equations of motion are 
\begin{eqnarray}
&&\partial^{\mu}J^{0}_{\mu}+\widetilde{C}J^{2}_{\mu}J^{1,\mu}=0\,, \qquad 
\partial^{\mu}J^{1}_{\mu}=0\,, 
\qquad 
\partial^{\mu}J^{2}_{\mu}+\widetilde{C}J^{0}_{\mu}J^{1,\mu}=0\,. 
\label{eom-wAdS}
\end{eqnarray}
By performing a double Wick rotation to the Lax pair in the squashed S$^3$ case 
\cite{FR}\footnote{We work on the notation used in \cite{KY3}. The convention of left and right in \cite{FR} is opposite to us.}, 
the Lax pair in the warped AdS$_3$ case is obtained as    
\begin{eqnarray}
&&L^{R}_{t}(x;\lambda)=\frac{1}{2}\left[L^{R}_{+}(x;\lambda)+L^{R}_{-}(x;\lambda)\right]\,, \qquad 
L^{R}_{x}(x;\lambda)=\frac{1}{2}\left[L^{R}_{+}(x;\lambda)-L^{R}_{-}(x;\lambda)\right]\,, \nonumber \\
&&L^{R}_{\pm}(x;\lambda)=-\frac{\sinh{\alpha}}{\sinh{\left(\alpha\pm\lambda\right)}}\left[-T^{0}J^{0}_{\pm}
+T^{2}J^{2}_{\pm}+\frac{\cosh{\left(\alpha\pm\lambda\right)}}{\cosh{\alpha}}T^{1}J^{1}_{\pm}\right]\,, 
\label{space_lax} 
\\
&&J_{\pm}=J_{t}\pm J_{x}\,,\qquad \widetilde{C}=\tanh^{2}{\alpha}\,. \nonumber 
\end{eqnarray}
The following commutation relation
\begin{eqnarray}
\left[\partial_{t}-L^{R}_{t}(x;\lambda),\partial_{x}-L^{R}_{x}(x;\lambda)\right]=0
\end{eqnarray}
leads to the equations of motion (\ref{eom-wAdS}).

\medskip 

The next task is to perform the scaling limit to the Lax pair (\ref{space_lax}). 
Let us first rewrite the Lax pair (\ref{space_lax}) by using $T^{\pm}$ and $J^{\pm}$ as 
\begin{eqnarray}
L^{R}_{\pm}(x;\lambda)=-\frac{\sinh{\alpha}}{\sinh{\left(\alpha\pm\lambda\right)}}\!\!\!\!&&
\left[-\frac{1}{2}\left(T^{+}+T^{-}\right)\left(J^{+}_{\pm}+J^{-}_{\pm}\right)+T^{2}J^{2}_{\pm}\right. \nonumber \\
&&\qquad \left.+\frac{\cosh{\left(\alpha\pm\lambda\right)}}{2\cosh{\alpha}}
\left(T^{+}-T^{-}\right)\left(J^{+}_{\pm}-J^{-}_{\pm}\right) \right]\,.
\end{eqnarray}
Consider the redefinition (\ref{lim}). 
$J^{\pm}_{\mu}=2{\rm Tr}(T^{\pm}J_{\mu})$ is also transformed under the redefinition as: 
\begin{eqnarray}
J^{-}_{\mu}\rightarrow\sqrt{\frac{2C}{\widetilde{C}}}J^{-}_{\mu}\,,\qquad 
J^{+}_{\mu}\rightarrow\sqrt{\frac{\widetilde{C}}{2C}}J^{+}_{\mu}\,.
\end{eqnarray}
When rescaling as $\lambda\rightarrow\alpha\lambda$\,, the Lax pair has the following form,  
\begin{eqnarray}
&& L^{R}_{\pm}(x;\lambda)=-\frac{\sinh{\alpha}}{\sinh{\left[\alpha\left(1\pm\lambda\right)\right]}} \nonumber \\
&&\times\left(
-T^{+}\left[\frac{1}{2}\left(1+\frac{\cosh{\left[\alpha\left(1\pm\lambda\right)\right]}}{\cosh{\alpha}}\right)J^{-}_{\pm}
+\frac{\tanh^{2}{\alpha}}{4C}\left(1-\frac{\cosh{\left[\alpha\left(1\pm\lambda\right)\right]}}{\cosh{\alpha}}\right)
J^{+}_{\pm}\right] \right. \nonumber \\
&&\qquad -T^{-}\left[\frac{1}{2}\left(1+\frac{\cosh{\left[\alpha\left(1\pm\lambda\right)\right]}}{\cosh{\alpha}}\right)J^{+}_{\pm}
+\frac{C}{\tanh^{2}{\alpha}}\left(1-\frac{\cosh{\left[\alpha\left(1\pm\lambda\right)\right]}}{\cosh{\alpha}}\right)
J^{-}_{\pm}\right]  \nonumber \\
&&\qquad 
+T^{2}J^{2}_{\pm}\biggr)\,.
\end{eqnarray}
Taking a limit in which $\alpha\rightarrow 0$ with $C$ and $\lambda$ fixed, the Lax pair in the right description 
of \sh sigma models 
is obtained as 
\begin{eqnarray}
L^{R}_{\pm}=-\frac{1}{1\pm\lambda}\left(-T^{+}J^{-}_{\pm}-T^{-}\left[J^{+}_{\pm} 
\mp C\left(\lambda\pm\frac{\lambda^{2}}{2}\right)J^{-}_{\pm}\right]+T^{2}J^{2}_{\pm}\right)\,.
\end{eqnarray}
Note that the $\alpha\rightarrow 0$ limit is the same as $\widetilde{C}\rightarrow 0$ 
because $\widetilde{C}=\tanh^{2}{\alpha}$\,.

\end{document}